\documentclass[letterpaper, 10 pt, conference]{ieeeconf}  

\IEEEoverridecommandlockouts                              

\usepackage{amsmath} 
\usepackage{amssymb}  
\usepackage[pdftex]{graphicx}
\usepackage[T1]{fontenc}
\usepackage[english]{babel}
\usepackage{tikz}
\usepackage{booktabs}
\usepackage{pgf}
\usepackage{import}
\usepackage{footnote}
\usepackage{subfig}
\usepackage{hyperref}
\usepackage{url}
\usepackage[style=ieee, dashed=false, sorting=none, doi=false, isbn=false, maxbibnames=6]{biblatex}
\title{\LARGE \bf
Systematic Categorization of Influencing Factors on Radar-Based Perception to Facilitate Complex Real-World Data Evaluation
}

\author{Maike Scholtes$^{1}$ and Lutz Eckstein$^{2}$
\thanks{*The research leading to these results is funded by the German Federal Ministry for Economic Affairs and Energy within the project `VVM - Verification \& Validation Methods for Automated Vehicles Level 4 and 5'.}
\thanks{$^{1}$Maike Scholtes is with the automated driving department of the Institute for Automotive Engineering (ika), RWTH Aachen University, 52074
Aachen, Germany
        {\tt\small maike.scholtes@ika.rwth-aachen.de}}
\thanks{$^{2}$Lutz Eckstein is head of the Institute for Automotive Engineering (ika), RWTH Aachen University, 52074 Aachen, Germany,
        {\tt\small lutz.eckstein@ika.rwth-aachen.de}}%
}

\begin{document}

\maketitle
\thispagestyle{empty}
\pagestyle{empty}

\begin{abstract}

For the assessment of machine perception for automated driving it is important to understand the influence of certain environment factors on the sensors used. Especially when investigating large amounts of real-world data to find and understand perception uncertainties, a smart concept is needed to structure and categorize such complex data depending on the level of detail desired for the investigation. Information on performance limitation causes can support realistic sensor modeling, help determining scenarios containing shortcomings of sensors and above all is essential to reach perception safety. The paper at hand looks into influencing factors on radar sensors. It utilizes the fact that radar sensors have been used in vehicles for several decades already. Therefore, previous findings on influencing factors can be used as a starting point when assessing radar-based perception for driver assistance systems and automated driving functions. On top of the literature review on environment factors influencing radar sensors, the paper introduces a modular structuring concept for such that can facilitate real-world data analysis by categorizing the factors possibly leading to performance limitations into different independent clusters in order to reduce the level of detail in complex real-world environments.

\end{abstract}

\section{INTRODUCTION}
\label{sec:Intro}
In the context of automated driving (AD), the assessment of machine perception increases in importance. However, sensors are highly sensible to the environment.
Therefore, real-world data of vehicles operating in open context (unstructured, public real-world environment, \cite{POD19}) is very complex to analyze. When seeking perception uncertainties \cite{DIE16} and possible environment factors causing shortcomings, it might be advisable to use already known factors from Advanced Driver Assistance Systems (ADAS) development of the past as a starting point.

This paper will provide an overview on influencing factors on radar perception used in automotive functions. The paper purposely does not use the term error, but describes possible limitations that can be seen when utilizing radar sensors. Those limitations cannot only be caused by an error of some sort, but also by physical capabilities and conditions imposed by the working principle of the sensor. For instance, an undetected object outside the sensor range would not be classified as an error. This paper analyzes the conditions that can possibly lead to performance limitations of radar-based perception on the basis of a review of different sources. Those sources can be clustered into: Expert knowledge \& literature, data driven \& experience based, customer \& market based and norms \& standards. The derived influencing factors are clustered according to the established 6-Layer Model (6LM) \cite{SCH20} and an additional \textit{Sensor Level}. This enhancement is made to, e.g., cover sensor specifications. The influencing factors are structured in this systematic manner, because they are highly dependent on different aspects of the environment and strong focusing is needed. This facilitates real-word data evaluation, as the complex domain can then be divided by only looking at certain layers, relevant for the studied environment factor, while neglecting others to reduce the level of detail.  

\section{STATE OF THE ART}

\label{sec:SateOfArt}

While this paper reviews multiple sources each looking into specific aspects of possible limitations of radar sensors, the authors are not aware of any work that provides such a joined, holistic overview. Furthermore, the results of the overview are structured in accordance with an adapted version of the 6LM, which was developed in current research projects. To the best of the authors knowledge this model has so far not been established in such a way in the context of machine perception for automated driving functions (ADF).

\subsection{Radar Sensor Application in Vehicles}
The application of radar sensors in vehicles has been researched for decades with first road tests aiming for suitability for series production in the 70's \cite{MEI14}. They increased massively in their importance with the introduction of first Adaptive Cruise Control (ACC) systems in the 90's. Radar sensors are standard for today's Forward Collision Warning (FCW), Autonomous Emergency Braking (AEB), ACC and blind spot warnings. With increasing development in the area of AD, radar sensors and also their further development (imaging radar, object classification etc.) have grown in importance again as they can generate an understanding of the environment. Radar sensors use electromagnetic waves in the radio domain that are emitted by the sensor, reflected by objects and received again by the sensor. In such a way, radar sensors are capable of determining the radial distance and angle to an object through the time-of-flight principle. Furthermore, they use the Doppler effect to determine the relative velocity of a detected object.

\subsection{Machine Perception}
In order to determine and cluster possible influencing factors on radar sensors, it is important to understand the different processing steps of such a sensor to get from sensing to actual perception. This paper will not look into any antenna characteristics or raw data in the sense of peaks in the frequency spectrum. We start on the application level. Nevertheless, multiple steps remain to get to an object list provided by an automotive sensor. The processing starts with determining detections, i.e. single points. In this work, the term detections does already include the identification of radial distance, azimuth and relative velocity (cf. \cite{WIN15}). The resolution of today's sensors has increased so that one object can trigger several detections. Therefore, those are clustered if they belong to the same object. Clustered detections provide information on objects in a particular scene (without time component). In order to increase the output quality of the sensor, perception algorithms take multiple time steps into account and track a potential object through several time steps. Such tracking includes prediction and association \cite{WIN15}. Based on previous measurements, possible future states are predicted. Those are then associated with the current measurement that can subsequently be assigned to an existing track.

Perception uncertainties are classified into three classes~\cite{DIE16}. \textit{State uncertainties} occur in case physical measured values, such as position or velocity, deviate from the ground truth; \textit{Existence uncertainties} occur if an object is detected that is not there (false positive (FP)) or if an object is not detected, even though it is present (false negative (FN)); \textit{Class uncertainties} state that an object is assigned to a wrong class, e.g. pedestrian instead of vehicle.

\subsection{Assessment of Automated Driving}
Purely statistical approaches for validation of ADF, where vehicles need to drive very long distances, have shown to be insufficient \cite{WAC15}. Therefore, the scenario-based validation process is favored and investigated in projects of the PEGASUS project family. In this approach test cases that challenge the ADF are generated. However, those approaches, so far, focus on the planning and acting part of the sense-plan-act architecture and mostly do not take sensing insufficiency into account. The assessment of machine perception introduces new challenges to the field, as, for instance, the open question of which manifestation of perception uncertainties should be classified as an error and what should be classified as relevant remains.

\section{6-LAYER MODEL FOR ENVIRONMENT DESCRIPTION AND NEEDED EXTENSIONS}
\label{sec:6LM}
\begin{figure}[t]
\centerline{\includegraphics[scale=0.5]{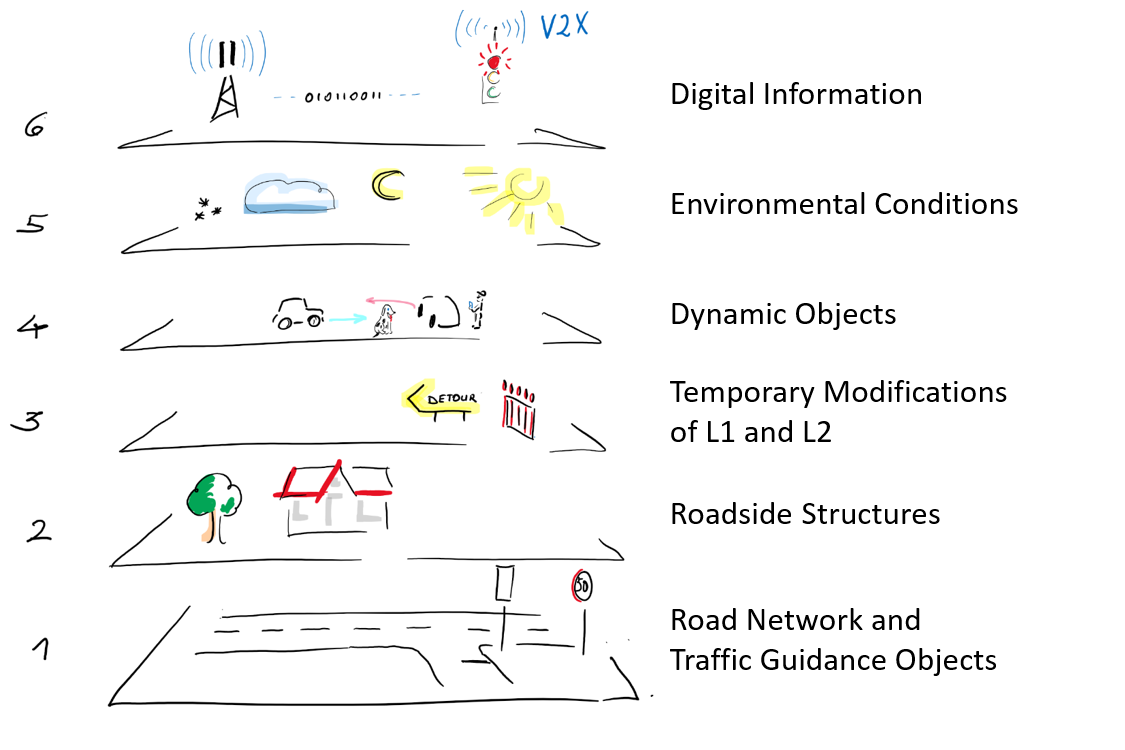}}
\caption{6LM for structured environment description.}
\label{fig:Layermodel}
\end{figure}

The 6LM is used to categorize the environment in order to provide a structured basis for complex real-world design domains. Its latest development can be found in \cite{SCH20}. An overview of the 6LM and its layers is revealed in Fig.~\ref{fig:Layermodel}.

\begin{figure}[t]
\centerline{\includegraphics[scale=0.4]{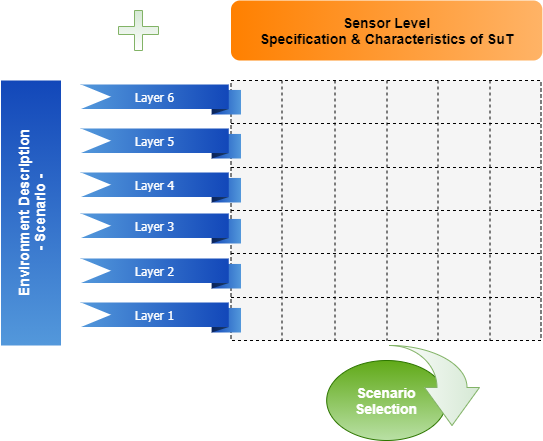}}
\caption{Structuring concept. Vertical structure for general environment description provided by the 6LM independent of the SUT. The additionally introduced \textit{Sensor Level} is presented horizontally and represents the specific SUT. The joined structuring concept can facilitate the selection of relevant scenarios for perception assessment.}
\label{fig:StructureOverview}
\end{figure}

It is important to note that the 6LM features an ``as it is'' description that is actor independent. This means that, e.g. occlusions from the point of view of one road user are not contained in the model, but can be calculated from the information given. Therefore, the idea of clustering possible influencing factors on radar-based perception by the different layers, requires that some additional matters aside the 6LM are addressed. First of all, the information given in the 6LM needs to be utilized to create an ego perspective. Since the 6LM is a holistic, actor independent description, this is perfectly possible with the given information. Once a given scenario is transferred into an ego perspective, the actual field of view (FoV) of the vehicle equipped with the sensor can be examined. Furthermore, sensor specific criteria need to be known in addition to the information given in the 6LM. Since the sensor itself is part of a vehicle (Layer~4) or the infrastructure for V2X concepts (Layer~6), it can be understood as being an element of this layer. However, the description of individual sensor specifications is not part of an holistic scenario description, but needed additionally when deriving relevant test cases for a specific setup / system-under-test (SUT) on the basis of the general scenario description and the specific subject to be tested. Therefore, sensor specifications are grouped in a separate additional description (the \textit{Sensor Level}) leaving the 6LM unaltered (cf. Fig. \ref{fig:StructureOverview}). Such an add-on for sensor specifications should contain information on the range, opening angle (azimuth and elevation), resolution / separation capability, mounting position, working principle, latency, software requirements etc. of the sensor in use.

Only the transformation into the ego perspective and the information on sensor specifications, such as technically achievable range, along with the information given in the 6LM make it possible to evaluate certain performance limitation causes. For instance, in case the exact position of different objects in the FoV of the sensor needs to be known. The physical boundaries of the sensor also induce that some limitations, presented in the following section, are not quite an error, let alone a phenomenon, but are purely due to physical capability or part characteristics preventing the sensor from perceiving an object.

\section{STRUCTURED DESCRIPTION OF INFLUENCING FACTORS ON RADAR-BASED PERCEPTION POSSIBLY CAUSING PERFORMANCE LIMITATIONS} 
\label{sec:StructDescr}

Since research and application of radar sensors in the automotive contexts dates back several decades, there exists numerous literature and experience regarding their working principle and possible environment influences on the technology. Therefore, for this paper existing sources were clustered into the four different groups: Expert knowledge \& literature, data driven \& experience based (testing procedures and driving campaigns), customer \& market based (information on existing functions utilizing radar sensors) and norms \& standards, which are subsequently analyzed.


This paragraph shortly introduces the used sources. Along with scientific literature, analyzed sources regarding testing campaigns are EuroNCAP AEB tests \cite{EUR19} and tests conducted in 2018 investigating progress in AD \cite{EUR18}. A similar testing campaign was repeated in 2020 \cite{EUR20b} and conducted by AAA in the US \cite{AAA18, AAA20c}. The AD test campaigns tested SAE Level 2 systems available on the market. Furthermore, NCAP testing protocols for dynamic brake support in the US were analyzed \cite{NCA15}. Experience gained through real-world data is taken from the post-processing of the radar data recorded for a naturalistic driving dataset \cite{GOR15} and a driving campaign conducted by comma.ai \cite{COM19}, looking for incorrect AEB braking where humans did not intervene. Furthermore, a test catalog for ACC testing was analyzed \cite{ANK96}. Regarding ADAS functions on the market utilizing radar sensors, the functions and their corresponding manuals of Tesla \cite{TES20} and Volvo \cite{VOL20b, VOL20c} were exemplarily studied. In terms of norms and standards, ISO/PAS 21448 \cite{SOT19} was analyzed. Furthermore, the new UN ECE standard for automated lane keeping systems of SAE Level 3 was investigated \cite{UN20}. The latter offers more concrete information in terms of relevant/hazardous driving scenarios, while both are rather unspecific in terms of further important environment conditions and, in particular, in case of ISO/PAS 21448, often based on examples. 

\begin{figure*}[t]
\centering
\centerline{\includegraphics[width=\textwidth, scale=0.75]{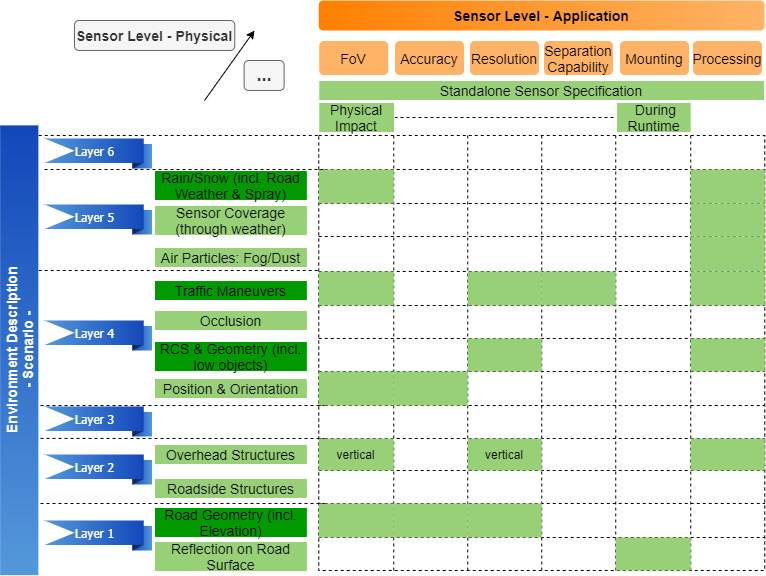}}
\caption{Overview of the different derived group of influencing factors and their assignment to different Layers of the 6LM and the \textit{Sensor Level}. Groups merged for visualization purposes are displayed in dark green color. The matrix structure visualizes the most important aspects of the working principle of the radar sensor responsible for the different influencing factors on the \textbf{application} level.}
\label{fig:Clustering1}
\end{figure*}

\begin{figure*}[t]
\centering
\centerline{\includegraphics[width=\textwidth, scale=0.75]{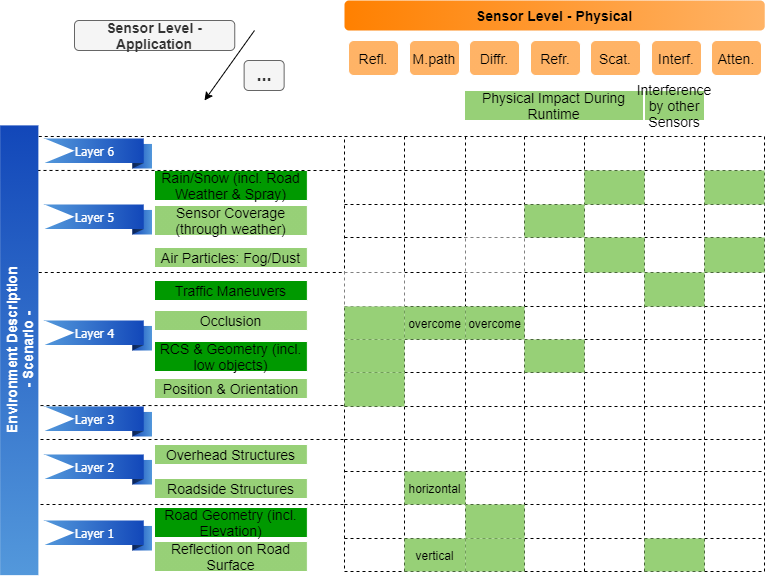}}
\caption{Similar overview than previous figure. However in this case, the matrix structure visualizes the most important aspects of the working principle of the radar sensor responsible for the different influencing factors on the \textbf{physical} level. The aspects considered are: Reflection, multipath reflections, diffraction, refraction, scattering, interference and attenuation.}
\label{fig:Clustering2}
\end{figure*}

In the following, aspects found through the analysis of the sources described above are clustered according to the 6LM and the \textit{Sensor Level} extensions formulated in Section \ref{sec:6LM}. Such clustering helps to structure and focus on different influencing factors on the individual layers possibly causing limitations of radar-based perception. Layer~3 is left out in the following analysis since it only contains temporary modifications of entities already discussed in Layer~1 and~2. An overview of the determined clusters is revealed in Fig.~\ref{fig:Clustering1} and Fig.~\ref{fig:Clustering2}. Additionally t providing an overview, the figures assign the relevant properties on the \textit{Sensor Level} responsible to for the influencing factors to the different groups. This assignment is made on the application layer as well as on the physical layer using the general properties of physical optics \cite{CAO17}.

\subsection{Sensor Level}
\label{subsec:SensorLayer}

\textbf{Standalone Sensor Specification.}
This cluster combines all properties that are caused by sensor specification, i.e., such that are not dynamically influenced through changing environment parameters. One major factor is the sensor's FoV. This includes range as well as opening angle (horizontally and vertically). The UN ECE norm \cite{UN20} also stresses the importance of a sufficient FoV by giving clear requirements on the longitudinal and lateral range of a sensor. Additionally, ISO/PAS 21448 \cite{SOT19} refers to range, resolution (horizontal and vertical), response time and durability as possible limitations. Objects that are not within the FoV of the sensor are by definition not directly detectable (with a few exceptions). However, the achievable FoV can also be reduced by various effects described below and, therefore, lead to missing detections. A factor influencing the FoV is the mounting position of the sensor. When choosing it too high, flatter objects might not be detected and measuring through windows becomes possible \cite{ANK96, VOL20c}. On the other hand, when mounting it too low, reflections on the road can lead to significant multipath reflections and interference disturbing the measurement \cite{WIN15, MUR13}. When integrating a sensor into a vehicle, other than the alignment of the sensor, the materials, possibly covering the sensor in its final mounting position, should be considered. Plastic and non-metallic paint usually do not influence the wave propagation (also depending on their thickness) \cite{WIN15, MUR13}, however, this needs to be considered by the manufacturer. Further aspects within this cluster are the resolution influencing the separation capability of a sensor. Moreover, so called packaging errors \cite{HAN16} caused by different processing steps can be included in the cluster at hand. Packaging errors contain influences induced by insufficient update rate or latency. Additionally, they include pure technical aspects such as maximum amount of objects a perception algorithm can process.

\textbf{Physical Impact During Runtime.}
Sensor positioning is a trade off between design decisions and technical requirements.  The positioning by the manufacturer should be chosen as best as possible taking this trade off into account. However, the sensor can also be affected during runtime and needs to be able to cope with this. E.g., vibrations and accidents can lead to misplacement and wrong calibration that need to be taken care of \cite{MUR13, CHO13}.
In \cite{CHO13} a recalibration method that uses the detections of the static environment is introduced.
Furthermore, damage to the sensor or covering of the sensor (for covering caused by weather conditions see Section \ref{subsec:groupLayer5}) can block the sensor or reduce its range \cite{TES20, VOL20c}. In the recently conducted tests by EuroNCAP \cite{EUR20b} reaction of the system to sensor blockage was tested.

\textbf{Interference Through Other Sensors.}
This cluster can be seen in combination with Layer~4 (and possible Layer~6 if V2X systems are installed) as it looks into influences by other radar sensors integrated into the ego vehicle (if vehicle features a multi sensor setup) or into other vehicles (or, in case of Layer~6, infrastructure). Interference from other vehicles is also stated as a possible sensor disturbance in ISO/PAS 21448 \cite{SOT19}. 
While \cite{WIN15} concluded that interference by other radar sensors most likely would not lead to FPs, the sensors could influence each other and reduce the sensitivity and, therefore, the detection capability.
In \cite{ANK96} this is also stated as a possible factor, for the use case highway, especially for traffic in road works where oncoming traffic is present. With increasing market penetration of radar sensors, vehicles being equipped with multiple sensors looking into different directions \cite{DIC16} and use in urban environments, where vehicles remain longer in the proximity of each other and face various directions, this problem might increase \cite{WIN15, BLO12, DIC15b}.

\subsection{Layer 1}
\label{subsec:groupLayer1}

\textbf{Elevation Profile.}
This cluster correlates with the Layer~4 cluster for ego driving maneuvers as change in elevation profile and deceleration or acceleration maneuvers can cause a change in pitch angle. A road featuring slopes and unevenness can influence the detection capability of a sensor. A change in elevation in front of the ego vehicle can lead to the target being outside of the FoV. This has, e.g., been observed in \cite{GOR15}. Furthermore, FPs could be created in case of dips in the road surface as the ego vehicle is approaching the surface \cite{ANK96}. An indication that this is a possible influencing factor is also given by EuroNCAP testing conditions requiring that the surface needs to be level and should not feature irregularities \cite{EUR19}.

\textbf{Road Geometry.} 
The geometry of the road can influence the detection capability of a sensor as the curvature in combination with the FoV of a sensor can lead to a vehicle not being detected or to the wrong vehicle being identified as the vehicle in front \cite{WIN15}.
While EuroNCAP AEB tests are conducted on straight roads \cite{EUR19}, the importance of the effect of curvature can be seen in various other testing campaigns such as \cite{EUR20b, ANK96} and also in the UN ECE norm \cite{UN20} that specifically requires testing of the scenarios on different road geometries. Furthermore, some vehicle manufactures' manuals also indicate that their systems should not be used in case of winding roads \cite{TES20, VOL20b}. 

\textbf{Reflections on the Road Surface.}
Additional to the above described dips, crests and unevenness the road surface can have, it can also feature so called micro and macro roughness. Effects of the road texture on reflections on the road surface are examined in \cite{DIE13} and \cite{SCH98}. It was found that with small incidence angles of the rays the micro roughness can be neglected and unevenness has a greater impact on the reflection as it increases attenuation. Furthermore, crests can lead to defocusing of the radar signal, while dips lead to a focusing. In general, reflections on the road surface can have positive as well as negative effects. Negative effects through reflections on the road surface in combination with a low mounting position of the sensor have been discussed in Section \ref{subsec:SensorLayer}. Whereas reflections on the road surface also cause multipath effects (see also Section \ref{subsec:groupLayer2}), those effects can be used in a positive way, too. In \cite{SCH98} it was shown that they can help to overcome occlusions as through reflections on road surface and underbody the radar sensor can see occluded objects in front of other vehicles. Furthermore, this effect allows for the calculation of orientation and size of an object \cite{SCH98, ROO16}. However, it can also lead to wrong distance measurements, if reflections from the underbody are detected instead of from the rear of the vehicle \cite{MUR13}. Additionally diffraction of waves on crests can enable the sensor ``to look through them''  \cite{SCH98}.

\subsection{Layer 2}
\label{subsec:groupLayer2}

\textbf{Overhead Structures.}
The correct processing of data in case of overhead structures, such as bridges and sign gantries, can be challenging for radar sensors. Please note that this cluster interacts with Layer~1 as traffic signs are described in this layer. However, since the information that a traffic sign is involved is, in this case, not as important as the presence of the overhead construction, it is placed in Layer~2. Furthermore, this cluster thematically interacts with Layer~4 \textit{Stationary Objects} (cf. Section \ref{subsec:groupLayer4}) as the occurring processing challenges correlate. Radar sensors feature a rather small vertical opening angle \cite{WIN15}. This complicates correct height identification of an object. Therefore, determining if a vehicle could drive underneath an object or if this object is a stationary obstacle is challenging. In \cite{DIE13}, an approach is shown using interference patterns in order to determine the height of an object. In case of objects located high above the surface, this pattern shows a high-frequent oscillation. The problem of classifying overhead structures as objects on the road can also be seen in real-world data; for instance, in the test data recorded by comma.ai, two out of ten breaking events of the AEB without intervention of the human driver were triggered by overhead constructions \cite{COM19}. Furthermore, a survey done with Tesla customers shows that the system causes some so called ``phantom braking events'', amongst others induced by such overhead structures \cite{TOM19}. ISO/PAS 21448 \cite{SOT19} also adds a scenario with a metal bridge to the test plan for perception system verification.

\textbf{Roadside Structures.}
An additional cause for FPs that can lead to unnecessary braking events is multipath propagation. In case of a multipath event, the emitted signal is not reflected straight off the object back to the sensor, but is scattered and reflected back to the sensor on a different path. The scattering can take place on many surfaces. The reflection on road surface and also the possible positive effects this can have, overcoming occlusion, has already been discussed in the Section \ref{subsec:groupLayer1}. Additionally, \cite{BAR12} showed a way to detect occluded pedestrians in between parked cars using multipath reflections.
Further structures that are often responsible for multipath effects in road traffic scenarios are such that belong to the roadside equipment. This includes large traffic signs, (tunnel) walls and guard rails.
Multipath reflections can, however, also occur on other surfaces such as other road users \cite{HOL19b} (cf. occluded pedestrian detection above). For this publication the influencing factor is, however, placed into the cluster at hand, as extended Layer~2 elements alongside of the road show great influence \cite{MUR13, VIS17}. FPs can, in particular, be caused by expanded objects (guard rails, concrete barriers etc.) as those induce false detections over multiple time steps that can not be eliminated by signal processing steps. The possible influence roadside equipment can have on radar sensors can also be concluded from the fact that for EuroNCAP AEB tests it is required that an area of three meters to each side of the test vehicle as well as the area 30 meters ahead is clear \cite{EUR19}.
Furthermore, FPs can be caused by steady objects such as tunnel walls or poles of guard rails. Especially the latter have a very high radar cross sections (RCS) (cf Layer~4 \textit{RCS and Geometry}) \cite{WIN15} and, therefore, generate many detections \cite{MEI17}. Since the object is steady, it is detected over several time steps and can be interpreted as an object moving along and with the same velocity as the ego vehicle \cite{ANK96}.

As stated above, roadside structures, especially guard rails, but also other stationary elements, can lead to numerous detections. This can be a problem since stationary objects beside the road could be interpreted as stationary road users as they might not be told apart. However, distinctive detections from safety structures can also help to determine the driving area \cite{DIC15b, LUN09}. In a first approach, when identifying the road area through detections on safety structures parallel to the course of the road, all objects not within the driving area could be neglected to overcome the confusion of stationary objects in general and stationary obstacles.

\subsection{Layer 4}
\label{subsec:groupLayer4}

\textbf{RCS and Geometry.}
Objects with different RCS and geometries can be a major influencing factor. The RCS of trucks, vehicles, motorcycles, bicycles and pedestrian is very different \cite{WIN15}, with two wheelers and pedestrians having a much smaller RCS, making it harder to detect them. While many official testing procedures do not address this as they only test with cars, the UN ECE norm \cite{UN20} clearly specifies scenarios with motorcycles and pedestrians as targets. Furthermore, analog to results of \cite{ANK96}, it also considers multiple consecutive targets with different RCS that might block each other (could also be triggered by highly reflective object such as guard rail). The effect of a highly reflective object triggering false detections can also be caused through access ladders on trucks. Those can act as a triple reflector \cite{WIN15}. User manuals of ADAS \cite{TES20, VOL20b} also clearly state that the systems do not or might not detect smaller objects such as motorcycles or pedestrians.

Furthermore, the geometry of objects can influence whether it is detected or not. Depending on the mounting position of the sensor, flat objects such as trailers or the back of towing trucks might not be detected \cite{AAA18, VOL20c}. This can also lead to the detected reflection point jumping from the back of the car to the driver's cabin which would cause a false distance measuring \cite{ANK96}.

\textbf{Position and Orientation.}
The position and orientation of an object within the FoV can influence the detection capability for different reasons. The distance between object and sensor has an influence on the strength of the signal received \cite{WIN15}. Furthermore, the RCS of an object is depended on its angle to the sensor, i.e., the orientation of the object. In case of a vehicle, the highest RCS can be achieved when looking straight at the front or back of the vehicle or when seeing it from the side, for other angles the RCS decreases \cite{SCH05}. The influence of different positions of the object in the FoV can also be seen in test protocols. EuroNCAP AEB \cite{EUR19} introduced tests with different degrees of overlap between ego vehicle and target. Furthermore, the UN ECE norm \cite{UN20} prescribes test where the target is purposely positioned at the side.

\textbf{Stationary Objects.}
The detection of stationary objects and the challenges arising in the processing of such data is crucial for radar sensors. The issue of a radar sensor allegedly not being able to detect stationary objects is related to the topic of overhead structures (Section \ref{subsec:groupLayer2}). A radar sensor is well capable of detecting stationary objects, however, those detections are often neglected (especially if not resulting from an object that has been moving at a previous timestep) in order to not trigger any false braking maneuvers, because overhead structures are falsely categorized as obstacles \cite{WIN15}. The difficulty of radar sensors correctly identifying stationary objects and reacting to it can be found in all clusters of data sources analyzed for this paper. All testing strategies include tests where the vehicle is approaching a stationary vehicle from behind \cite{EUR19, EUR18, EUR20b, AAA20c, NCA15, ANK96} and/or cut-out tests where a faster vehicle cuts out in between the ego vehicle and a stationary vehicle enabling the view onto the stationary object \cite{EUR18, EUR20b, AAA18, UN20}. Also manuals for ADAS show that the manufacturer are aware of such problems. The Volvo manual \cite{VOL20b} excludes slow or standing vehicles from the detection capability of its function. Tesla changed its manual from stating that it is unable to detect stationary vehicles to a more specific description that this can especially be the case for velocities above 80~kph and in a constellation that is similar to the above described cut-out scenario \cite{TES20}. Furthermore, the survey done with Tesla customers indicates that so called ``failure to brake'' amongst others occurred for stationary vehicles \cite{TOM19}. The results of EuroNCAP and AAA testing campaigns in 2018 and 2020 \cite{EUR18, EUR20b, AAA18, AAA20c} also indicate that this is an existing problem, especially for higher velocities.

\textbf{Occlusion.}
Missing detections due to occlusion are not caused by errors in the processing of radar data. On the contrary, the working principle can even help to overcome it (cf. \textit{Reflection on Road Surface} in Section \ref{subsec:groupLayer1} and multipath propagation in Section \ref{subsec:groupLayer2}). However, occlusion needs to be listed as a factor in order to be complete. It can be caused by various objects, also such of Layer~1 (e.g larger traffic signs) and Layer~2 (e.g. buildings or trees). It is, however, listed here, in Layer~4, as Layer~4 objects are especially important when analyzing occlusions in the road area.

\textbf{Low Objects Driven Over.}
Smaller objects that can be driven over, especially when made out of metal, can have an influence on the radar sensor. Those objects could also be objects assigned to Layer~1, such as, e.g., manhole covers or marking nails. However, they are placed into Layer~4 here. Prominent examples for Layer~4 objects that can lead to FP and false braking maneuvers are e.g. coke cans \cite{WIN15}. Also during the drives conducted by comma.ai two unnecessary brakings were caused by such objects on the the road: a plastic container and a metal plate \cite{COM19}. Testing for FPs due to metal plates, is also required by NCAP \cite{NCA15}. However, \cite{SCH98} has shown that at least small objects right above the ground, like screws, will not cause any problems as they can not be detected. Furthermore, \cite{SCH98} introduced a similar procedure as \cite{DIE13} in order to detect if an object is flat and can be driven over.

\textbf{Dynamic Maneuvers of Target and Ego Vehicle Maneuvers.}
Dynamic maneuvers of the target such as braking, turning or weaving within the lane \cite{AAA18} can cause problems in the processing of the sensor data in terms of tracking and prediction. The prediction of the target location and the actual target location can then deviate and cause FNs and FPs, respectively, as seen in \cite{GOR15}. Furthermore, dynamic maneuvers of the ego vehicle itself can influence the perception. While turning has an influence on the FoV and is also a challenge for processing algorithms \cite{ANK96}, braking or acceleration maneuvers can lead to large pitch angles and cause similar effects than slopes do (cf. Section \ref{subsec:groupLayer1}).

\textbf{Other Traffic Scenarios.}
Most test maneuvers are rear traffic maneuvers. Testing in oncoming traffic or cross traffic is rare or non existent. Recently EuroNCAP included a turn maneuver through oncoming traffic \cite{EUR19}. Oncoming traffic is also often excluded from the design domain of current ADAS \cite{VOL20b}, but it will need further investigation in the future when radar technology is utilized for ADF. The same holds for cross traffic maneuvers that are not addressed in testing campaigns today. In \cite{DIC16} and \cite{DIC15a} first requirements for handling intersections and roundabouts have already been identified.
Additionally, an object driving very closely to the ego vehicle with a small offset could be undetected in case of a sensor featuring focused rays \cite{ANK96}. This is comparable to a cut-in scenario where a vehicle cuts in in front of the ego vehicle and might be undetected depending on the sensors FoV \cite{ANK96, VOL20c}. Cut-ins can be problematic and are, therefore, included in the testing specifications of UN ECE \cite{UN20} and were also tested in other testing campaigns by EuroNCAP \cite{EUR18, EUR20b}. 
Further traffic constellations that might be problematic are vehicles driving next to each other with nearly the same velocity, as those could be detected as one single vehicle \cite{ANK96}. Also vehicles driving in the neighboring lane close to the own lane could falsely be detected with a wrong azimuth angle and interpreted as in front of the ego vehicle \cite{UN20}. Similarly, the passage between two vehicles can be problematic \cite{ANK96}.

\subsection{Layer 5}
\label{subsec:groupLayer5}

\textbf{Environmental Conditions.} Influences on automotive perception sensors caused by environmental conditions are widely discussed. Even though, radar is known to be a sensing technology that is not very prone to such factors, effects can still be observed. Radar sensors can be influenced by \textbf{rain and snow} in terms of attenuation and backscattering. With increasing rain intensity, measured RCS can be reduced by a significant amount \cite{HAS16}. This leads to a reduced range during intensive rain \cite{ZAN19}. However, not only the rain intensity is important, but also the droplet size and falling direction. If the size is in the same order of magnitude as the wavelength of the sensor, scattering can occur \cite{WIN15, ANK96, ZAN19}. In general, ice has a smaller influence than water. Therefore, the influence of snow (dependent on its water content) is not as high as the one caused by rain \cite{HAS06}. The same holds for \textbf{sensors covered by water} or ice. While a sensor being covered by water can lead to unwanted focusing of the rays \cite{WIN15}, an ice cover has no impact \cite{ARA06}. 
Other particles in the air, such as during fog or dust, show little to no impact on radar sensors \cite{HAS17, RYD09}. Furthermore, a water cover on the road can lead to \textbf{spray} that can possibly cause FPs \cite{WIN15}. With the test constellation chosen in \cite{SHE98}, however, it was found that spray does mostly contain smaller water droplets that do not cause backscattering. Furthermore, no attenuation effect that could prevent detections was found. However, this is most likely also largely dependent on the amount of water on the road and the tire and chassis' shape of the vehicle causing the spray.

\section{DISCUSSION AND OUTLOOK}
\label{sec:discussion}

The paper provides a structured and sound list of environment factors possibly imposing performance limitations on radar-based perception systems. It reviews results from research on automotive radar sensors and structures it according to the 6LM to reduce the complexity. In this paper, only influencing factors on existing radar sensors known in literature and applied in the field are analyzed. With future developments in radar sensing, such as better object classification, imaging radar or joint radar and communication systems, the list of factors will increase and with the latter also include Layer~6 of the 6LM. Furthermore, additional requirements such as an increased FoV \cite{DIC16} will be introduced when utilizing radar technology for ADFs.
When evaluating real-world perception data to find perception uncertainties, the developed concept can be used to keep the level of detail as low as possible. This is achieved by concentrating on different layers of the 6LM plus the \textit{Sensor Level} individually and neglecting other layers, depending on the different factors investigated. Future work, that is planned, will apply this concept to real-world datasets and elaborate on needed parameters in order to retrieve perception uncertainties caused by the presented factors. Furthermore, it will formulate requirements on datasets in order to be able to investigate the limitations caused by some of the derived factors. The discovered environment factors leading to perception uncertainties will help to identify events for the sense part of the sense-plan-act architecture that possibly lead to hazardous behavior of an ADF. Therefore, this will support identification of challenging scenarios when assessing perception safety.
Since the paper focuses on radar-based perception to provide a starting point when investigating perception data, future research can also investigate the transferability of the concepts to other perception systems utilizing different sensors or a combination of such.






\printbibliography

\end{document}